\documentclass[12pt]{article}
\usepackage{float}
\usepackage{amssymb,amsmath,amsfonts,latexsym,graphicx}
\usepackage[numbers]{natbib}
\usepackage{subcaption}
\usepackage{placeins}

\begin{document}

\begin{center}
\Large \bf CapsuleNet: A Deep Learning Model to Classify GI Diseases using EfficientNet-b7 \rm

\vspace{1cm}


\large Aniket Das$\,^a$, \large  Ayushman Singh$\,^a$, \large  Nishant$\,^a$, \large  Sharad Prakash$\,^a$

\vspace{0.5cm}

\normalsize


$^a$ Indian Institute of Information Technology, Ranchi

\vspace{5mm}


Email: {\tt aniket.2022ug1052@iiitranchi.ac.in},
{\tt ayushman.2022ug3011@iiitranchi.ac.in},
{\tt nishant.2022ug3010@iiitranchi.ac.in},
{\tt sharad.2022ug3017@iiitranchi.ac.in}

\vspace{1cm}

\end{center}

\abstract{Gastrointestinal (GI) diseases represent a significant global health concern, with Capsule Endoscopy (CE) offering a non-invasive method for diagnosis by capturing a large number of GI tract images. However, the sheer volume of video frames necessitates automated analysis to reduce the workload on doctors and increase the diagnostic accuracy. In this paper, we present CapsuleNet, a deep learning model developed for the Capsule Vision 2024 Challenge, aimed at classifying 10 distinct GI abnormalities. Using a highly imbalanced dataset, we implemented various data augmentation strategies, reducing the data imbalance to a manageable level. Our model leverages a pretrained EfficientNet-b7 backbone, tuned with additional layers for classification and optimized with PReLU activation functions. The model demonstrated superior performance on validation data, achieving a micro accuracy of 84.5\% and outperforming the VGG16 baseline across most classes. Despite these advances, challenges remain in classifying certain abnormalities, such as Erythema. Our findings suggest that CNN-based models like CapsuleNet can provide an efficient solution for GI tract disease classification, particularly when inference time is a critical factor.}

\section{Introduction}\label{sec1}
Gastrointestinal (GI) diseases are a growing global health concern, with Capsule Endoscopy being a key tool for non-invasive diagnosis. This enables direct visualization of the GI tract through small cameras but results in a large amount of video frames which needs to be manually reviewed. This is not only time-consuming but is also error-prone. Hence, Automating this process using AI can significantly reduce the burden on doctors and improve diagnostic accuracy.

The Capsule Vision 2024 Challenge aims to advance AI models for multi-class abnormality classification in Vision Capsule Endoscopy(VCE) images. Organized by MIAAI, MISAHUB, and CVIP 2024, the challenge provides a dataset covering 10 distinct GI abnormalities, such as Angioectasia, bleeding, and polyps. In this paper, we present our approach to addressing this challenge, focusing on data preprocessing and cleaning, model development, and performance evaluation for GI tract disease classification.
\section{Methods}\label{sec2}
The training dataset \cite{Handa2024} that we received was highly imbalanced, with the largest class (normal images) having 28000 instances while the smallest class (worms) had just 158 images. This called for a number of data augmentation and pre-processing techniques before training.
\subsection{Data Augmentation}
Given the large number of images required for training deep learning models, we decided to use 5000 images per class, by using various augmentation methods. However, despite the large augmentation methods used, the smallest class still had minimal variance, causing the model to severely overfit, resulting in poor results on the test dataset.
To alleviate this issue, we decided to decrease the number of augmented images to 1500 per class. In classes with more than 1500  instances, we randomly sampled the images, whereas, for classes with fewer images, we augmented them to 1500 samples. The augmentations used were: Random Rotation, Horizontal and Vertical flipping, Elastic Transforms, Gaussian noise and blurring, and random brightness contrast.
\subsection{Image Preprocessing}
We used standard preprocessing steps of image resizing, BGR to RGB conversion, and normalization. We also used Contrast limited adaptive histogram equalization (CLAHE), but this contributed to the overfitting of the models, and was removed from the pipeline.
\subsection{Architecture}

\begin{figure}[h]
    \centering
    \includegraphics[width=0.2\linewidth ]{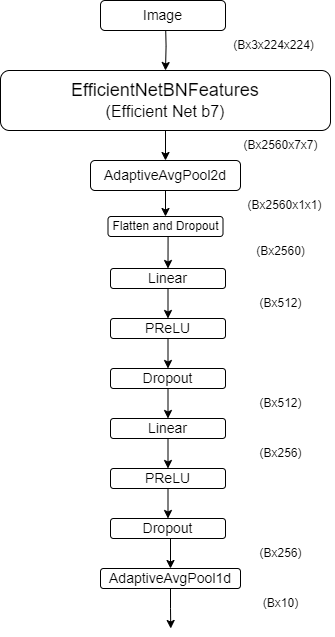}
    \caption{Block diagram of the developed pipeline.}
    \label{fig:CapsuleNet}
\end{figure}During initial literature exploration, we encountered project Monai \cite{monaieffnet} and decided to used its pretrained models for our classification task. Among the various tested models, including Densenet, Resnet, VGG and EfficientNet, we found that EfficientNet-b7 gave us best validation results. Hence, our final pipeline is based on a monai efficientnet backbone, and some additional layers for final classification. We tested the classifier with ReLU, Leaky ReLU, and PReLU activations and found PReLU activations to marginally outperform the other two activations.

\FloatBarrier
\section{Results}\label{sec3}

\begin{figure}[H]
\centering
\begin{subfigure}{.5\textwidth}
  \centering
    \includegraphics[width=\linewidth]{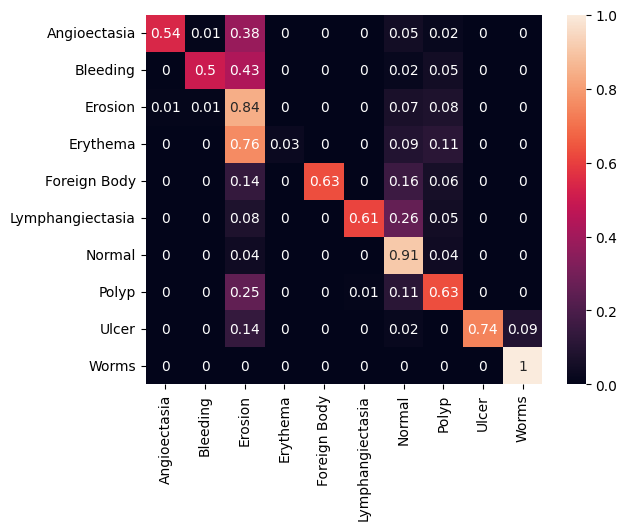}
    \caption{Confusion Matrix for Val dataset (\textit{\% of $<y-label>$ classified as $<x-label>$})}
    \label{fig:enter-label}
\end{subfigure}%
\begin{subfigure}{.5\textwidth}
  \centering
  \includegraphics[width=0.6\linewidth]{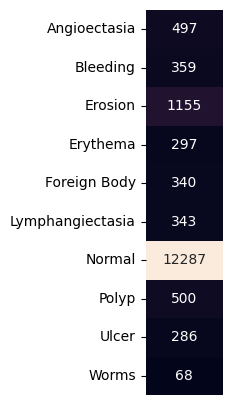}
    \caption{Support}
    \label{fig:enter-label}
\end{subfigure}
\caption{Validation Metrics}
\label{fig:test}
\end{figure}
\begin{table}[h]
\begin{tabular}{|l|lll|lll|l|}
\hline
                          & \multicolumn{3}{c|}{\textbf{VGG16 (by MISAHUB)}}         & \multicolumn{3}{c|}{\textbf{CapsuleNet}}                 &                  \\ \hline
\textbf{Class}            & \textbf{precision} & \textbf{recall} & \textbf{f1-score} & \textbf{precision} & \textbf{recall} & \textbf{f1-score} & \textbf{support} \\ \hline
\textbf{Angioectasia}     & 0.33               & 0.50            & 0.40              & 0.88               & 0.54            & 0.67              & 497              \\
\textbf{Bleeding}         & 0.51               & 0.57            & 0.54              & 0.84               & 0.50            & 0.62              & 359              \\
\textbf{Erosion}          & 0.29               & 0.40            & 0.33              & 0.43               & 0.84            & 0.57              & 1155             \\
\textbf{Erythema}         & 0.13               & 0.37            & 0.19              & 0.91               & 0.03            & 0.06              & 297              \\
\textbf{Foreign Body}     & 0.33               & 0.67            & 0.44              & 0.90               & 0.63            & 0.74              & 340              \\
\textbf{Lymphangiectasia} & 0.37               & 0.51            & 0.43              & 0.83               & 0.61            & 0.70              & 343              \\
\textbf{Normal}           & 0.96               & 0.78            & 0.86              & 0.97               & 0.91            & 0.94              & 12287            \\
\textbf{Polyp}            & 0.21               & 0.38            & 0.26              & 0.32               & 0.63            & 0.43              & 500              \\
\textbf{Ulcer}            & 0.48               & 0.81            & 0.61              & 0.99               & 0.74            & 0.85              & 286              \\
\textbf{Worms}            & 0.60               & 0.69            & 0.64              & 0.71               & 1.00            & 0.83              & 68               \\ \hline
\textbf{Accuracy}         &                    &                 & 0.71              &                    &                 & 0.85              & 16132            \\
\textbf{macro avg}        & 0.42               & 0.56            & 0.47              & 0.78               & 0.64            & 0.64              & 16132            \\
\textbf{weighted avg}     & 0.81               & 0.71            & 0.75              & 0.90               & 0.85            & 0.85              & 16132            \\ \hline
\end{tabular}
\caption{Result Comparisons}
\end{table}

On the Validation set, we achieved a micro accuracy value of  0.845 and a macro accuracy of 0.643. The f1-score achieved on individual classes outperformed the VGG16 baseline model provided by Misahub on all the classes except Erythema, with an overall accuracy of 0.85 compared to the baseline accuracy of 0.71. The model performed poorly for the erythema class and confused it with erosion throughout the validation process. 

\FloatBarrier
\section{Discussion}\label{sec4}

\begin{figure}[h]
\centering
\begin{subfigure}{.5\textwidth}
  \centering
  \includegraphics[width=\linewidth]{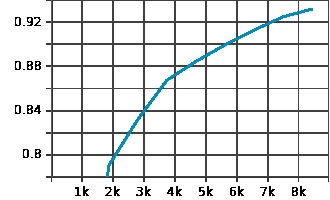}
  \caption{Macro Accuracy}
  \label{fig:sub1}
\end{subfigure}%
\begin{subfigure}{.5\textwidth}
  \centering
  \includegraphics[width=\linewidth]{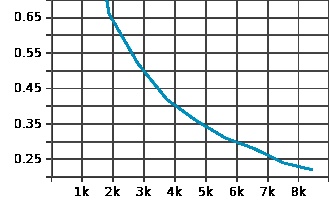}
  \caption{Loss}
  \label{fig:sub2}
\end{subfigure}
\caption{Training Metrics}
\label{fig:test}
\end{figure}

\begin{figure}[h]
\centering
\begin{subfigure}{.5\textwidth}
  \centering
  \includegraphics[width=\linewidth]{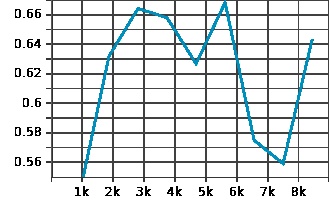}
  \caption{Macro Accuracy}
  \label{fig:sub1}
\end{subfigure}%
\begin{subfigure}{.5\textwidth}
  \centering
  \includegraphics[width=\linewidth]{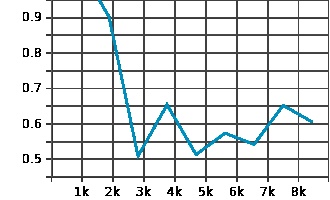}
  \caption{Loss}
  \label{fig:sub2}
\end{subfigure}
\caption{Validation Metrics}
\label{fig:test}
\end{figure}
Here are the training and validation metrics, with the number of training steps on x-axis and the corresponding metric on y-axis.
We compared multiple models to the baseline and adjusted the loss functions and hyperparameters as needed. One of the methods that we initially employed to tackle the imbalanced classes was using Focal loss instead of the standard Cross Entropy loss. The model did show some promise initially but worsened significantly after just a few epochs and was discarded.\\
After our first augmentation (5000 images per class) resulted in poor results, we augmented each class to 8 times the number of instances in that class (capping the maximum value to 5000) and used class weights for each class to overcome the class imbalance, but could not converge to acceptable weight values, as the class weights kept interfering with the model training and resulted in a non-converging loss and a poorly trained classifier.\\
The final augmentation gave us 1500 instances per class and the model was trained on this data with Cross Entropy loss function, and no class weights. 
\FloatBarrier
\section{Conclusion}\label{sec5}
In this paper we present CNN based model for VCE image classification of 10 classes. Our proposed architecture is based on the efficientnet model from monai framework, and a simple classifier layer. This simple architecture gave us acceptable results on the validation set, while keeping the inference time to a minimum. We believe that CNN based models would be more suitable towards Medical Image Classification tasks such as VCE, where a reduced inference time significant.
\FloatBarrier
\section{Acknowledgments}\label{sec6}
As participants in the Capsule Vision 2024 Challenge, we fully comply with the competition's rules as outlined in \cite{handa2024capsule}. Our AI model development is based exclusively on the datasets provided in the official release in \cite{Handa2024}.

\bibliographystyle{unsrtnat}
\bibliography{sample}

\end{document}